\documentstyle[axodraw,graphicx]{article}

\def \eslt{\not\!\!{E_T}}

\begin{document}
\vspace*{-1in}
\renewcommand{\thefootnote}{\fnsymbol{footnote}}
\begin{flushright}
UH-511-1096-06\\
\end{flushright}
\vskip 5pt
\begin{center}
{\Large{\bf WW Fusion in Higgsless Models}}
\vskip 25pt
{\sf Rahul Malhotra}
\vskip 10pt
{\small High Energy Physics Group, University of Hawaii at Manoa} \\

\vskip 20pt

{\bf Abstract}
\end{center}

\begin{quotation}
{\small Recently several Higgsless models of electroweak symmetry breaking have been
proposed in which unitarity of $W,Z$ scattering amplitudes is
partially restored through a tower of massive vector gauge bosons.
These massive states are expected to couple mainly to $W$ and $Z$ and should
appear as resonances in $WW$ and $WZ$ fusion at the CERN Large
Hadron Collider (LHC). We study the LHC discovery reach for the
first neutral state, $V_1^0$, through the reaction
$W^+W^- \rightarrow V_1^0 \rightarrow W^+W^- \rightarrow e^\pm \mu^\mp
+ \eslt$. The background from $t\bar{t}$, $\tau^+\tau^-$ and $W$ pair
production is calculated and dual forward jet tagging as well as a
mini-jet veto in the central detector region are applied. The maximal 
$P_T(e,\mu)$
distribution is found to have a fat tail for large $P_T$ that rises above
the backgrounds and allows $5\sigma$ discovery for $V_1^0$ masses above
$1$ TeV and total integrated luminosity of $300$ fb$^{-1}$.}
\end{quotation}

\vskip 20pt

\setcounter{footnote}{0}
\renewcommand{\thefootnote}{\arabic{footnote}}
%\vfill
%\clearpage
%\setcounter{page}{1}
%\pagestyle{plain}

%% INTRODUCTION
  \section{Introduction}
Though the Standard Model (SM) has been very successful in
explaining a range of observations at hadron colliders and LEP, the
exact mechanism for electroweak symmetry breaking (EWSB) remains a
mystery. However, in most models of new physics, a Higgs scalar is
assumed to be responsible for EWSB as it also preserves the
unitarity of $WW$ and $WZ$ scattering amplitudes
\cite{higgsunitarity1,higgsunitarity2}.

Recently some alternative methods for EWSB have been proposed in
models containing more than 4 dimensions \cite{higgsless} and in the
so-called deconstructed theories \cite{deconstructed}. In the
extra-dimensional models, the boundary conditions on fields in the 
extra dimensions result
in constraints on the masses and ordering of Kaluza Klein (KK) modes
such that the lowest set of modes looks like the Standard Model. At
the same time, violation of unitarity in $WW$ and $WZ$ scattering is
delayed to $\Lambda \sim 5-10$ TeV due to exchange of a KK tower of
massive vector bosons. We denote these bosons by $V_N^0$ and
$V_N^\pm$; where $N = 0,1,2,...$ represents the KK level of the
state. The $N=0$ modes are the familiar $W^\pm$ and $Z$ bosons.

Such Higgsless models initially faced serious challenges from
Precision Electroweak Constraints (PEC) as they predicted a positive
$S$ parameter \cite{spositive}. These issues were resolved recently
in warped space Higgsless models by fermion delocalization
\cite{szero} which also results in vanishing overlap between
wavefunctions of the higher ($N>0$) KK modes and those of the SM
fermions. Hence the primary couplings of the KK vector boson states
would be between themselves and the $W$, $Z$ particles. Other
remaining issues concern the top quark mass, which are currently
under active investigation \cite{topissues} and have been resolved
in some models.

As Higgsless theories have now matured into viable models of new
physics at the TeV scale, it is important to investigate their
phenomenological signatures at the LHC. In this paper we concern
ourselves with $WW$ scattering where the $W$s are radiated off
initial state quarks. Fig. 1(a-c) shows the $WW \to WW$ scattering
diagrams involving intermediate $V_N^0$ particles. Normally, without
a Higgs boson or the KK modes, the $WW \to WW$ scattering amplitude
has terms that grow as $E^2$ or $E^4$. If there is a Higgs boson,
then these terms are identically zero. However, in Higgsless
extra-dimensional models, the terms are cancelled by the KK modes
provided certain identities between their masses and couplings to $W,Z$ bosons 
are
satisfied. These are:
\begin{equation}
g_{WWWW} = g_{WWZ}^2 + g_{WW\gamma}^2 + \sum_{N=1}^{\infty}
(g_{WWV}^N)^2
\end{equation}
and
\begin{equation}
 4g_{WWWW}M_W^2 = 3g_{WWZ}^2 M_Z^2 +
3\sum_{N=1}^{\infty} (g_{WWV}^N M_N^0)^2
\end{equation}
where $g_{WWV}^N$ is the trilinear $WWV_N^0$ coupling and $M_N^0$ is
the mass of $V_N^0$. As shown in \cite{matchev}, considerations of
series convergence as well as a survey of Higgsless models show that
the sum rules are nearly saturated by the first $N=1$ mode. In that
case Eqns. (1,2) reduce to
\begin{equation}
g_{WWV}^1 \approx  g_{WWZ} \frac{M_Z}{\sqrt{3}M_1^0}
\end{equation}
which gives us the approximate coupling of the first mode to $W$
bosons. From Fig. 1(c) it is clear that the $V_N^0$ should show up
as resonances in pair production of $W$s. Also, the final state
decay rate $V_N^0 \to WW$ should be close to $100\%$, assuming 
the $V_N^0$ decouple from SM fermions.

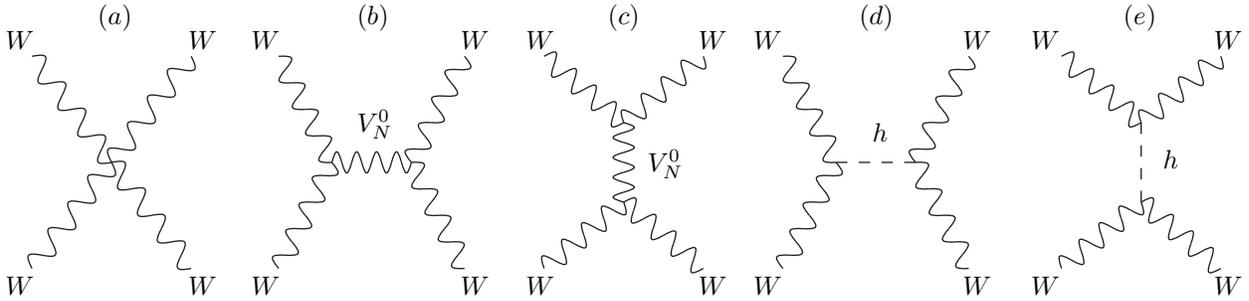
\begin{figure}[t]
\begin{center}
\begin{picture}(500,100)(0,0)

\Photon(10,10)(70,90){4}{8}
\Photon(10,90)(70,10){4}{8}
\Text(0,0)[lb]{$W$}
\Text(0,100)[lt]{$W$}
\Text(80,100)[rt]{$W$}
\Text(80,0)[rb]{$W$}
\Text(35,110)[lt]{$(a)$}

\Photon(103,10)(123,50){4}{4}
\Photon(103,90)(123,50){4}{4}
\Photon(123,50)(153,50){4}{4}
\Photon(153,50)(173,90){4}{4}
\Photon(153,50)(173,10){4}{4}
\Text(93,0)[lb]{$W$}
\Text(93,100)[lt]{$W$}
\Text(183,100)[rt]{$W$}
\Text(183,0)[rb]{$W$}
\Text(133,70)[lt]{$V_N^0$}
\Text(133,110)[lt]{$(b)$}

\Photon(203,10)(233,35){4}{4}
\Photon(233,35)(233,65){4}{4}
\Photon(203,90)(233,65){4}{4}
\Photon(263,90)(233,65){4}{4}
\Photon(263,10)(233,35){4}{4}
\Text(197,0)[lb]{$W$}
\Text(197,100)[lt]{$W$}
\Text(273,100)[rt]{$W$}
\Text(273,0)[rb]{$W$}
\Text(243,55)[lt]{$V_N^0$}
\Text(228,110)[lt]{$(c)$}

\Photon(293,10)(313,50){4}{4}
\Photon(293,90)(313,50){4}{4}
\DashLine(313,50)(343,50){4}
\Photon(343,50)(363,90){4}{4}
\Photon(343,50)(363,10){4}{4}
\Text(283,0)[lb]{$W$}
\Text(283,100)[lt]{$W$}
\Text(373,100)[rt]{$W$}
\Text(373,0)[rb]{$W$}
\Text(328,65)[lt]{$h$}
\Text(323,110)[lt]{$(d)$}

\Photon(398,10)(428,35){4}{4}
\DashLine(428,35)(428,65){4}
\Photon(398,90)(428,65){4}{4}
\Photon(458,90)(428,65){4}{4}
\Photon(458,10)(428,35){4}{4}
\Text(388,0)[lb]{$W$}
\Text(388,100)[lt]{$W$}
\Text(468,100)[rt]{$W$}
\Text(468,0)[rb]{$W$}
\Text(438,55)[lt]{$h$}
\Text(423,110)[lt]{$(e)$}

\end{picture}
\end{center}
\caption{Feynman diagrams for $WW \to WW$ scattering. In Higgsless
models we only have diagrams (a-c). In the Standard Model however,
all diagrams contribute, with the $V_N^0$ replaced by the $Z$ boson
only.}
\end{figure}

A similar resonance in $WZ \to WZ$ scattering was exploited in
\cite{matchev} to show that the lightest $V_1^\pm$ particles can be
discovered at the LHC with masses up to $1$ TeV using about $60$
fb$^{-1}$ of integrated luminosity. The purely leptonic decay of the
final state $WZ$ allows for a clean observation of the first
$V_1^\pm$ state, including mass reconstruction. The rest of the KK
modes are not expected to be observable due to their heavier masses
and smaller couplings to $WZ$.

In the case of $V_1^0$ however, the $WW + 2$ forward jets final
state is much more difficult to separate from the background,
especially given the approximately $800$ pb of $t\bar{t}$ production
cross-section expected at the LHC \cite{topcross}. Mass reconstruction is only
possible in the $WW \to 2j + l\nu$ or $WW \to 4j$ channels, which
results in a minimum of $4$ jets in the final state once the forward
jets are counted. We concentrate on the much cleaner leptonic
channel $WW \to e\mu + \eslt$ where mass reconstruction is not
possible but appropriate cuts can be made to reveal an excess.

To enhance the signal significance we employ the technique of
minijet veto which has been shown to be promising for the case of
intermediate and heavy Higgs production via $W$ fusion i.e. the
Higgs discovery reaction $WW \to H \to WW \to e\mu + \eslt$ \cite{higgssearch}.

%% PHYSICS BACKGROUND
\section{Physics Background}

The dominant physics backgrounds to the final state of $2$ "forward
jets" + $e^\pm \mu^\mp + \eslt$ come from $pp \to t\bar{t} + X$
production, followed by $pp \to WW + 2j$. We define ``forward jets'' 
as having transverse momentum $P_T > 20$ GeV and pseudorapidity $2 <
|\eta| < 4$, with one jet having positive and the other negative
$\eta$.

\subsection{$t\bar{t}$ background}

The main physics background to the $e^\pm \mu^\mp + \eslt$ signal
arises from $t\bar{t}$+jets production, due to its nearly 800 pb of
cross section at the LHC. Top pair production at the LHC is strongly
dominated by the gluon fusion channel, $gg \to t\bar{t}$, while the
$q\bar{q} \to t\bar{t}$ channel is less than 10\% of the total.
Emission of additional quarks or gluons leads to $t\bar{t}+j$ and
$t\bar{t}+2j$ events. 

If no additional partons are emitted, the $b$
quarks from the decaying tops are required to be in the forward jet
configuration. Then, the $t\bar{t}+j$ cross-section, where the $b$
quarks are required to be forward, allows us to estimate the minijet
activity in the central region. To estimate minijet activity in
$t\bar{t}+j$, for the configuration where the final-state light
quark or gluon, and one of the $b$ quarks are identified as forward
tagging jets, we calculate $t\bar{t}+2j$. Lastly, we also calculate
$pp \to t\bar{t} + 2j$, where the final state light quarks or gluons
are the two tagging jets and the $b$ quarks from top decay must pass
the minijet cuts (next section). Overall the methodology is very
similar to that used in \cite{higgssearch}. The matrix elements for
all three processes were obtained using CompHEP. The narrow-width
approximation is used for top and $W$ decays. A $K$ factor of 2 is
chosen which normalizes total $t\bar{t}$ production to 800 pb for
our renormalization and factorization scale $\mu_R=\mu_F=2M_{top}$.

\subsection{$WW+2j$ background}

This background consists of (a) radiation of $W^+,W^-$ during quark
anti-quark scattering and (b) QCD corrections to $W^+W^-$
production. Normally, (b) is higher than (a), but (a) includes $WW$
fusion diagrams, which give final states very similar to the signal.
Therefore, in the final analysis, (a) turns out to be larger than
(b). We also calculate the $WW+3j$ cross section to estimate minijet
activity. A similar factorization scale is chosen as for the signal
process $V^\pm+2j$, which is the invariant mass of the final state
$W$s i.e. $\mu_R=\mu_F=M_{WW}$. Variation of scale by a factor of 2
results in uncertainties as high as 30\%, which means that in the
absence of NLO calculations, experimental calibration might be
necessary. CompHEP is used here also to generate matrix elements.

\subsection{$\tau^+ \tau^- + 2j$ background}

The $e\mu+2j$ final state can be obtained from $\tau\tau+2j$
production due to the leptonic decay of $\tau$ leptons. We use the
collinear approximation for $\tau$ decay \cite{collinear}. However,
the $e,\mu$ from $\tau$ decay are comparatively soft and we see later that 
in most cases this
background turns out to be insignificant ($< 1$\%) in comparison to
the two described above.

\section{Analysis}

We define "forward jets" as final state partons satisfying the
following selection criterion
\begin{equation}
P_T > 20 GeV; 2 < |\eta| < 4
\end{equation}
Furthermore, the two forward jets need to be on opposite ends of the
beam-line. The leptons are required to satisfy
\begin{equation}
P_T > 20 GeV; |\eta| < 2.5
\end{equation}
as well as isolation from the jets
\begin{equation}
\Delta R_{j,l} > 0.7
\end{equation}
to suppress the background from semi-leptonic decays of QCD $b$
quark production. With these cuts, the two types of $WW+2j$
backgrounds are already comparable in magnitude, which is due to the
fact that Standard Model $WW$ fusion processes (not including the
Higgs boson) pass the forward jet cut (4) naturally. This technique
of isolating $WW$ fusion processes is well established in the
literature \cite{forwardtagging1,forwardtagging2,forwardtagging3}.

The $W$ pairs from decay of the heavy $V_1^0$ are close to being 
back-to-back as the $V_1^0$ are slow-moving for higher masses. Also,
the $W$'s are boosted enough that the $e\mu$ pair from their decay
has a large separation angle i.e. $\cos\theta_{e\mu}$ is
predominantly negative. For the $t\bar{t}$ and the $WW$ backgrounds
however, the $\cos\theta_{e\mu}$ distribution is roughly uniform.
Therefore, we employ the following cut
\begin{equation}
\cos\theta_{e\mu} < 0.1
\end{equation}
which keeps $\sim 85$\% of the signal, but removes more than 50\% of
the background.

We use real $\tau$ reconstruction to reduce the $\tau\tau+2j$
background. In the collinear approximation, if $x_1,x_2$ are the
$\tau$ momentum fractions carried by $e,\mu$ respectively, then we
have the equations
\begin{equation}
(1-\frac{1}{x_1}){\bf P_T^1} + (1-\frac{1}{x_2}){\bf P_T^2} = {\bf
\eslt}
\end{equation}

\begin{figure*}[t!]
\centering
\begin{tabular}{|c||ccc|ccccccc||crc|}
    \hline
\textbf{\em Cuts}  & & $V_1^0+2j$ &   & &  $t\bar{t}+jets$ & & $\tau\tau+2j$ & & $WW+2j$ & & & $S/\sqrt{B}$ & \\
    \hline \hline
forward jets, leptons (4,5,6)       & & 0.742 & & & 1150 & & 17.1 & & 8.12 & & &0.375  &\\
+ $b$ veto                     & & -- & & & 74.8 & & -- & & -- & & & 1.29 &\\
+ $M_{JJ}$, $\cos\theta_{e\mu}$ (7,11) & & 0.551 & & & 19.5 & & 2.01 & & 2.98 & & & 1.93 &\\
+ real $\tau$ rejection (9,10)        & & 0.541 & & & 17.8 & & 0.290 & & 2.73 & & & 2.05 &\\
+ minijet veto                 & & 0.487 & & & 5.69 & & 0.0884 & & 1.37 & & & 3.15 &\\
+ $P_T^{max}(e,\mu) > 230$ GeV & & 0.138 & & & 0.0271 & & $\sim 10^{-4}$ & & 0.0583 & & & 8.12 &\\
    \hline
\end{tabular}
\caption{Signal and background cross-sections and significance for
$M_V = 700$ GeV for the
$pp$ collider at the LHC with center-of-mass energy $\sqrt{s} = 14$ TeV and a
total integrated luminosity of 300 fb$^{-1}$. The
effect of various cuts on the cross-sections is shown. All cross-sections
are in fb. Without the mini-jet veto, $S/\sqrt{B}$ is still $\sim 5.9$ for a $V_1^0$ of the same mass.}
\end{figure*}
where $P_T^{1,2}$ are the original transverse momenta vectors of the
$\tau$ leptons that give rise to $e,\mu$ respectively. These two
equations for $x,y$ coordinates can be solved simultaneously to
yield $x_{1,2}$. If the $e,\mu$ really did come from a pair of
$\tau$s, then it must be true that
\begin{equation}
0 < x_{1,2} < 1
\end{equation}
Furthermore, we can use the $x_{1,2}$ to reconstruct the invariant
mass of the $\tau$s, $M_{\tau\tau}$, and then veto events that fall
near the $Z$ pole as that is where most of the $\tau\tau$ pairs
originate. Finally, events that satisfy condition (9) and
\begin{equation}
M_Z - 30 GeV < M_{\tau\tau} < M_Z + 30 GeV
\end{equation}
are vetoed. This cuts down the $\tau\tau$ background by a factor of
7 while leaving the signal and other backgrounds almost untouched.

The most problematic background is the $W^+W^-$ pairs from
$t\bar{t}+X$ production. The first step in controlling this is a
veto on $b$ or $\bar{b}$ jets with $P_T > 20$ GeV in the region
between the two forward jets. It is to be noted that $b$ tagging is
not required here, just rejection of events with a central jet. This
by itself cuts down the $t\bar{t}$ background by a factor of 15. It
is still the largest background by far, but now of the same order of
magnitude as the other backgrounds.

Besides this, a cut on the invariant mass, $M_{JJ}$, of the two
forward jets
\begin{equation}
M_{JJ} > 650 GeV
\end{equation}
is useful in reducing the $t\bar{t}$ background by about 60\% with little 
effect on the signal.
This is because QCD processes at the LHC typically occur at lower
invariant masses than the signal. Similarly QCD corrections to $WW$
production are also reduced by about the same factor.

The minijet veto on jets with $P_T > 20$ GeV in the region between
the forward jets is applied, to match the $b$ veto condition above.
We follow the procedure given in \cite{higgssearch} which has been
shown to be useful for Higgs discovery in the $WW$ fusion channel
extensively in the literature \cite{higgssearch2,higgssearch3}. For
our cuts, this procedure shows a survival probability of 90\% for
signal events, but only 32\% for the $t\bar{t}$ background and 53\%
for the $WW+2j$ background.

In the $V_1^0 \to WW \to e\mu+\eslt$ decay channel, explicit
reconstruction of the invariant mass of the two $W$'s is not
possible as there are two neutrinos. Nor are the $W$'s boosted
enough to permit a collinear approximation for their decay products.
However, the maximal $P_T$ between the $e,\mu$ does have a peak at
an energy that increases monotonically with the mass of the $V_1^0$.
We do not employ a "window" cut though around the peak as there are
too few signal events. Instead, we note that the transverse momentum 
distribution of the lepton carrying the higher $P_T$, hereafter denoted by 
$P_T^{max}(e,\mu)$, has a fat tail towards higher momenta that rises above 
the background. We then demand that
$P_T^{max}(e,\mu)$ be greater than a lower threshold which is chosen
for each $M_V$ such that the signal significance is maximized.

We find that the optimal threshold for $P_T^{max}(e,\mu)$
varies roughly linearly between 100 and 350 GeV for $M_V$ between 400
and 1200 GeV. However, we also ask that there be at least 10 signal
events, which does not always allow this threshold to be used.

The step-by-step effect of the cuts is shown in Fig. (2).
 
\section{Results}

We calculate signal significance over the background as a
function of the mass of $V_1^0$ for integrated luminosities of
${\cal L} = 30, 100, 300$ fb$^{-1}$, as shown in Fig. (3). 
\begin{figure}[h!]
\begin{center}
\includegraphics[angle=-90]{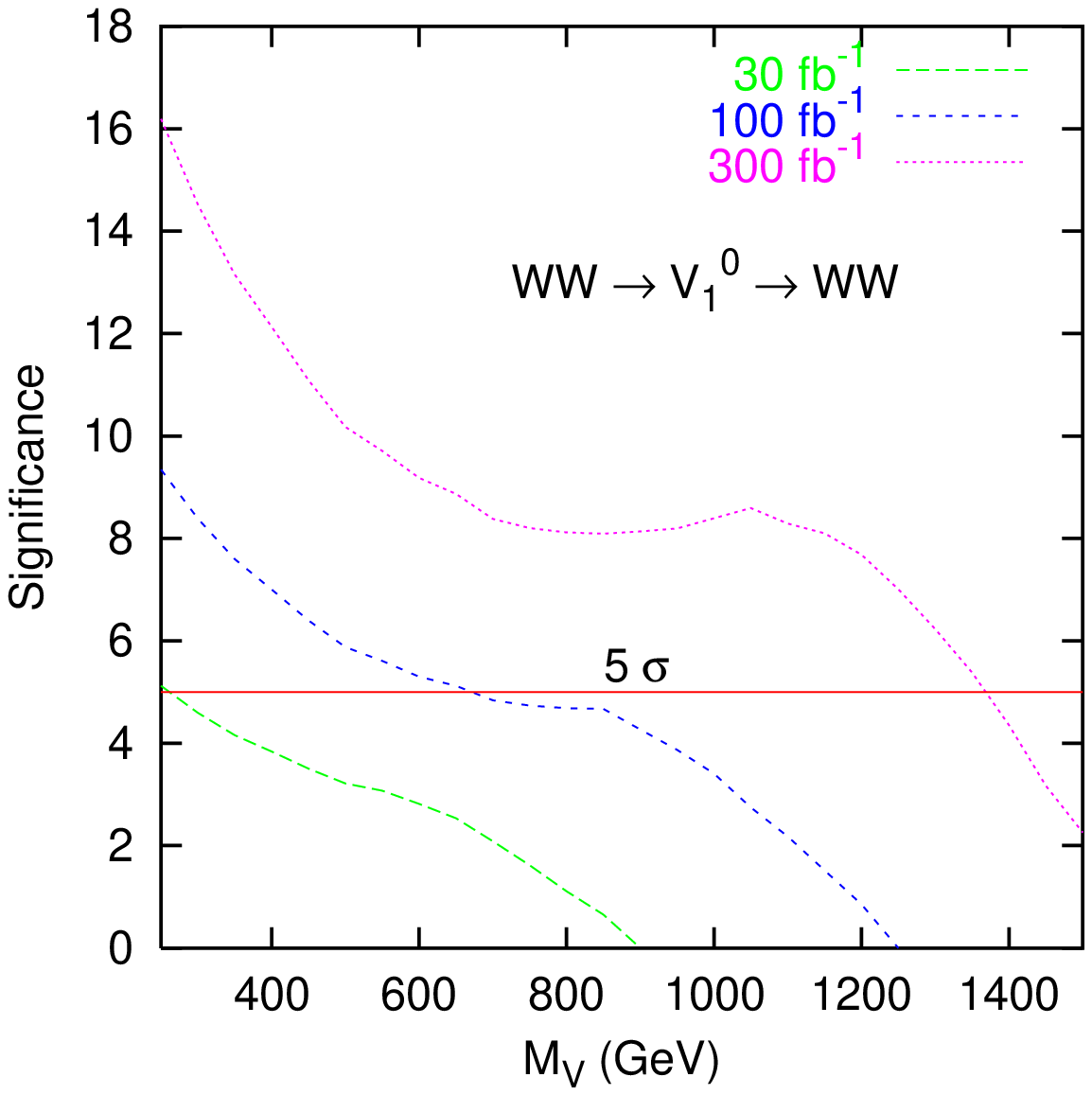}
\end{center}
\caption{LHC discovery potential for $V_1^0$, the first Kaluza-Klein 
excitation of the $Z^0$ boson in Higgsless models, at $\sqrt{s}$ = 
14 TeV.}
\end{figure}
For low
luminosity running (30 fb$^{-1}$), the $5\sigma$ discovery potential
is poor, with only masses below $\sim 300$ GeV being accessible. However,
for high luminosity running (300 fb$^{-1}$), $V_1^0$ as heavy as
1.35 TeV can be discovered. As shown in the figure, between
$M_V = 800-1000$ GeV the discovery potential actually improves for
this luminosity because more energetic $V_1^0 \to WW$ decays lead
to higher transverse momenta for the final state $e,\mu$.
This offsets
to some extent the effect of the lower $VWW$ coupling (which falls off
as $1/M_V$), and the more restricted phase space.

Our requirement that there be a minimum of 10 signal events implies
that for low luminosities we cannot always use the most optimal lower
cutoff on $P_T^{max}(e,\mu)$. Therefore, the signal significance drops
sharply beyond a point for all three luminosities considered.

A minijet veto could be challenging to realise. If this cut is not
used, the discovery potential is still quite substantial, upto $M_V
= 900$ GeV, for ${\cal L} = 300$ fb$^{-1}$.

It is clear therefore that both $V_1^\pm, V_1^0$ particles, which
are the key signatures of extra-dimensional (and deconstructed)
Higgsless models, are within reach of the LHC. While discovery of
the $V_1^\pm$ would provide a smoking gun \cite{matchev}, searches
for the $V_1^0$ would provide important confirmation, 
and can be conducted in the same channels as those for a
heavy Higgs.

%% ACKNOWLEDGEMENTS
\section{Acknowledgements}

I would like to thank D. Dicus, X. Tata, K. Melnikov and H. Paes for
useful discussions and help given throughout the course of this
work. This work was supported in part by the United States
Department of Energy under Contract No. DE-FG-03-94ER40833.


\begin{thebibliography}{99}

\bibitem{higgsunitarity1} D. A. Dicus and V. A. Mathur, Phys. Rev. D
{\bf 7}, 3111 (1973); C. H. Llewellyn Smith, Phys. Lett. B {\bf 46},
233 (1973); J. M. Cornwall, D. N. Levin and G. Tiktopoulos, Phys.
Rev. Lett. {\bf 30}, 1268 (1973) [Erratum-ibid. {\bf 31}, 572
(1973)]; Phys. Rev. D {\bf 10}, 1145 (1974) [Erratum-ibid. D {\bf
11}, 972 (1975)].

\bibitem{higgsunitarity2} B. W. Lee, C. Quigg and H. B. Thacker, Phys.
Rev. Lett. {\bf 38}, 883 (1977); Phys. Rev. D {\bf 16}, 1519 (1977);
M. S. Chanowitz and M. K. Gaillard, Nucl. Phys. B {\bf 261}, 379
(1985).

\bibitem{higgsless} C. Csaki, C. Grojean, H. Murayama, L. Pilo and J. Terning, Phys.
Rev. D {\bf 69}, 055006 (2004); C. Csaki, C. Grojean, L. Pilo and J.
Terning, Phys. Rev. Lett. {\bf 92}, 101802 (2004); Y. Nomura, JHEP
0311, 050 (2003); C. Csaki, C. Grojean, J. Hubisz, Y. Shirman and J.
Terning, Phys. Rev. D {\bf 70}, 015012 (2004).

\bibitem{deconstructed} R.Sekhar Chivukula, Elizabeth H. Simmons, Hong-Jian He,
Masafumi Kurachi and Masaharu Tanabashi (Tohoku U.), Phys. Rev. D
{\bf 71},115001 (2005); Hong-Jian He, Int. J. Mod. Phys. A {\bf 20},
3362 (2005); R. Casalbuoni, AIP Conference Proceedings {\bf 806},
104 (2006).

\bibitem{spositive} H. Davoudiasl, J. L. Hewett, B. Lillie and T. G. Rizzo,
Phys. Rev. D {\bf 70}, 015006 (2004); G. Burdman and Y. Nomura,
Phys. Rev. D {\bf 69}, 115013 (2004); H. Davoudiasl, J. L. Hewett,
B. Lillie and T. G. Rizzo, JHEP 0405, 015 (2004); J. L. Hewett, B.
Lillie and T. G. Rizzo, hep-ph/0407059; R. Barbieri, A. Pomarol, R.
Rattazzi and A. Strumia, hep-ph/0405040.

\bibitem{szero} Giacomo Cacciapaglia, Csaba Csaki, Christophe
Grojean and John Terning, Phys. Rev. D {\bf 71}, 035015 (2005)

\bibitem{topissues} Roshan Foadi and Carl Schmidt, Phys. Rev. D {\bf 73}, 075011
(2006); Giacomo Cacciapaglia, Csaba Csaki, Christophe Grojean,
Matthew Reece and John Terning, Phys. Rev. D {\bf 72}, 095018
(2005).

\bibitem{matchev} Andreas Birkedal, Konstantin Matchev and Maxim Perelstein,
Phys. Rev. Lett. {\bf 94}, 191803 (2005).

\bibitem{topcross} Stefano Catani, Michelangelo L. Mangano, Paolo Nason and Luca Trentadue,
Phys. Lett. B {\bf 378}, 329 (1996); B.W. Harris, E. Laenen, L.
Phaf, Z. Sullivan and S. Weinzierl, Phys. Rev. D {\bf 66}, 054024
(2002); ATLAS Technical Design Report
http://atlas.web.cern.ch/Atlas/GROUPS/PHYSICS/TDR/access.html.

\bibitem{higgssearch} David L. Rainwater and D. Zeppenfeld, Phys. Rev. D {\bf 60}, 113004
(1999)

\bibitem{collinear} K. Hagiwara, A. D. Martin, and D. Zeppenfeld, Phys. Lett. B {\bf 235}, 198 (1990)

\bibitem{forwardtagging1} R. N. Cahn, S.D. Ellis, R. Kleiss and W.J. Stirling,
Phys. Rev. D {\bf 35}, 1626 (1987); V. Barger, T. Han, and R. J. N.
Phillips, Phys. Rev. D {\bf 37}, 2005 (1988); R. Kleiss and W. J.
Stirling, Phys. Lett. B {\bf 200}, 193 (1988); D. Froideveaux, in
Proceedings of the ECFA Large Hadron Collider Workshop, Aachen,
Germany, 1990, edited by G. Jarlskog and D. Rein (CERN report 90-10,
Geneva, Switzerland, 1990), Vol II, p. 444; M. H. Seymour, ibid, p.
557; U. Baur and E. W. N. Glover, Nucl. Phys. B {\bf 347}, 12
(1990); Phys. Lett. B {\bf 252}, 683 (1990).

\bibitem{forwardtagging2} V. Barger, K. Cheung, T. Han, and R. J. N. Phillips,
Phys. Rev. D {\bf 42}, 3052 (1990); V. Barger et al., Phys. Rev. D
{\bf 44}, 1426 (1991); V. Barger, K. Cheung, T. Han, and D.
Zeppenfeld, Phys. Rev. D {\bf 44}, 2701 (1991); erratum Phys. Rev. D
{\bf 48}, 5444 (1993); Phys. Rev. D {\bf 48}, 5433 (1993); V. Barger
et al., Phys. Rev. D {\bf 46}, 2028 (1992).

\bibitem{forwardtagging3} D. Dicus, J. F. Gunion, and R. Vega, Phys. Lett. B {\bf 258},
475 (1991); D. Dicus, J. F. Gunion, L. H. Orr, and R. Vega, Nucl.
Phys. B {\bf 377}, 31 (1991).

\bibitem{higgssearch2} T. Plehn, David L. Rainwater and D. Zeppenfeld, Phys. Rev. D {\bf 61}, 093005
(2000); N. Kauer, T. Plehn, David L. Rainwater and D. Zeppenfeld,
Phys. Lett. B {\bf 503}, 113 (2001); Vernon D. Barger, R.J.N.
Phillips and D. Zeppenfeld, Phys. Lett. B {\bf 346}, 106 (1995);
Vernon D. Barger, King-man Cheung, Tao Han and D. Zeppenfeld, Phys.
Rev. D {\bf 44}, 2701 (1991); Erratum-ibid. D {\bf 48}, 5444 (1993).

\bibitem{higgssearch3} D.A. Dicus, J.F. Gunion, L.H. Orr and R. Vega, Nucl. Phys. B {\bf 377}, 31
(1992).

\end{thebibliography}
\end{document}